\documentclass[11pt,a4paper,oneside]{article}
\usepackage[latin1]{inputenc}
\usepackage{amsmath}
\usepackage{amsfonts}
\usepackage{amssymb}
\usepackage{url}
\usepackage{latexsym}
\usepackage{graphicx}
\usepackage[a4paper]{geometry}
\newcommand{\mma}[0]{Mathematica }

\date{}

\author{Reinhold Kainhofer \and Reinhard V. Simonovits}
\title{\bf \huge M@th Desktop and MD Tools\\[0.5em]
\Large\em Mathematics and Mathematica Made Easy for Students}

\makeatletter
\begin{document}
\maketitle
\begin{abstract}
We present two add-ons for \mma for teaching mathematics to undergraduate and high school students. These two applications, M@th Desktop (MD) and M@th Desktop Tools (MDTools), include several palettes and notebooks covering almost every field. The underlying didactic concept is so-called "blended learning", in which these tools are meant to be used as a complement to the professor or teacher rather than as a replacement, which other e-learning applications do. They enable students to avoid the usual problem of computer-based learning, namely that too large an amount of time is wasted struggling with computer and program errors instead of actually learning the mathematical concepts.

M@th Desktop Tools is palette-based and provides easily accessible and user-friendly templates for the most important functions in the fields of Analysis, Algebra, Linear Algebra and Statistics. M@th Desktop, in contrast, is a modern, interactive teaching and learning software package for mathematics classes. It is comprised of modules for Differentiation, Integration, and Statistics, and each module presents its topic with a combination of interactive notebooks and palettes.



\end{abstract}

\section{Introduction}
Computer-based mathematics courses have become more and more popular in recent years. Instead of having to do all calculations by hand, students are supported by a powerful computational tool, which enables the students themselves to focus on the actual mathematical concepts. As a consequence, various e-learning packages have been developed, and much literature on the theory of computer-based teaching has been published. However, computer-based mathematics teaching also has its difficulties, namely that the students often only learn how to enter the data into the program and request a certain result without actually learning what they (or  rather their machines) are doing and why they are doing it.

In this paper, we present M@th Desktop  Tools and M@th Desktop, both of which are Mathematica-based teaching and learning packages for mathematics which try to find a good balance between the different approaches to computer-based teaching. In Section~\ref{concept}, we will present the didactic concept behind the package, the so-called "blended learning" in which the lecturer and computer-based e-learning software complement each other. In Section~\ref{MDT} we will first present the palette-based M@th Desktop Tools package, which is a set of palettes which provide templates for the most important functions of several mathematical fields. M@th Desktop Tools is directed at  students and undergraduates who need to use the computer, and Mathematica in particular, to solve mathematical problems on their own. Finally, we present in Section~\ref{MD} the full-feature courseware package M@th Desktop, which provides notebooks introducing various mathematical concepts and gives exercises to reinforce those concepts once introduced. Palettes which again contain templates for the most important functions in each mathematical area are once again included.

\section{The didactic concept}\label{concept}
Both M@th Desktop and M@th Desktop Tools are based on the didactic concept of "blended learning". Like in e-learning, the students study the mathematical concepts with the use of a computer program (MD and MDTools in this case). Blended-learning tools, however, are not intended to replace a professor or teacher, but rather to complement him or her. Thus, blended learning is a combination of traditional teaching and e-learning. Ideally, the professor first explains the mathematical concepts in front of the whole class using traditional teaching methods. The students are then asked to perform calculations and solve exercise problems themselves using MD and MDTools. The interface of MD and MDTools is tailored to the needs of high school teachers and university lecturers giving undergraduate introductory courses in mathematics.

\subsection{Why teach mathematics using a computer?}
Most people will agree that the essential part of studying mathematics is to understand the underlying ideas and not to memorize how to solve one specific problem. Understanding the concept allows the students to apply their knowledge to other problems and combine ideas from several different problems. In conventional mathematics classes, however, students often do exactly that: They learn by heart each step in the solution of one specific problem. Unfortunately, not only do they forget this solution path very quickly, they also have to invest more time in learning all possible exercises rather than learning how to apply the general concept itself to any specific problem. Since some calculations, like solving a multidimensional linear algebra problem, are very time-consuming if done by hand, there is no obvious benefit to the students if they have to use the same algorithms (e.g. Gauss elimination in this case) over and over again. 

In contrast, these steps can be automated using a computer so that the students can concentrate on the essential concept of learning how to formulate a given problem in mathematical terms (equations). Of course, when the students first encounter a specific mathematical problem, they first need to learn and understand how to use the algorithms, but after they have, for example, solved several systems of linear equations by hand, there is no reason for them to continue doing the same mechanical calculation for every problem. In particular, when a system of linear equations is encountered as a side-problem in a problem that is really meant to demonstrate some other mathematical concept, its time-consuming solutiong by hand distracts the student's attention to some extent from the actual concept the problem is supposed to demonstrate.

This point is essential: The computer is simply intended to remove from the student the burden of doing the same tedious calculations over and over again once they really do understand how to do it by hand. This is also one of the  main points of the theory of blended learning: The teacher is still needed to give the students the mathematical foundation, only it is now the computer which will be doing the hard and boring work afterwards. Buchberger \cite{Buchberger:1989, Buchberger:1992} for example emphasizes the importance of this phase, which he calls the white-box phase. It is the time in which the lecturer needs to impart a well-founded knowledge of the mathematical concepts and also the mathematical details of the calculation itself, before the computer can be used for the calculations.

Then, however, the computer as a calculating tool allows the students to concentrate on the concepts rather than the details of the calculation.

\subsection{What distinguishes M@th Desktop from other e-learning software?}
Most e-learning tools cover a scope similar or even greater than that of MD and MDTools. 
However, they either provide very specialized sections for each mathematical topic or have a steep learning curve. As an example, let us again look at solving systems of linear equations. Many mathematical programs for students present them with the needed theory of the Gaussian algorithm (as text to be read by the students before they can actually do some work, thus this part tries to replace the teacher), and then allow them to do some exercises, but only with limited flexibility. 

Usually, in applications of the first type, the students are presented with a table for the system of linear equations where they have to enter the coefficients of the equation system, and the computer prints out the solution. 

Another category of e-learning software is general mathematics applications, where someone provides files that can be used like textbooks. The problem with these, however, is that the students need to learn how to use the mathematics application before they can use it to learn. Even after they know how to work with the application (i.e. how to define equations, how to solve them, etc), the probability that they will make a simple input error is considerable, with the consequence that much time is wasted in finding this error. In particular, the students in our example need to know how to define the equations exactly, and how to apply the solve function.

M@th Desktop and M@th Desktop Tools try to cure these two problems and combine their solutions. First, Mathematica allows a far greater flexibility than the programs described in the first category of e-learning software. Since MD and MDT are based on Mathematica, they inherit this flexibility. However,  this is not enough, as MD and MDT would still suffer from exactly the same problems described for the second category. To avoid these problems, MD and MDT provide the user with several specialized palettes containing templates for the most needed and important tasks. The user simply has to insert the relevant data specific to the problem. In contrast to the first category, however, the user can save every expression and work further with it.

\vspace{0.5em}

Another important issue which distinguishes MD and MDT from other e-learning software is the concept of blended learning as discussed above. The teacher or lecturer still fills the crucial role in the educational process while being supported by the software rather than being made obsolete by it.

\section{Didactic research on M@th Desktop}
Currently, M@th Desktop and MDTools are heavily used within the EU Project "Computer Algebra and the Web: Modern Tools for Understanding Mathematics." The program is being used for teaching at several high schools and Universities of Applied Sciences in Austria and Germany. 

M@th Desktop has been developed in cooperation with Univ. Prof. Dr. Bernd Thaller at the Institute of Mathematics of the University of Graz. Several diploma theses have been written \cite{Fink:2001,Siller:2002,Welik:2002}\footnote{For more information, see \url{http://www.uni-graz.at/imawww/diplomarbeiten/index.html}. All completed diploma theses about M@th Desktop can be downloaded there.} or are currently being written about the palettes and the didactical background of M@th Desktop.

\section{M@th Desktop Tools (MDTools)}\label{MDT}

\begin{figure}\centering
\begin{minipage}{4cm}
\includegraphics[width=4cm]{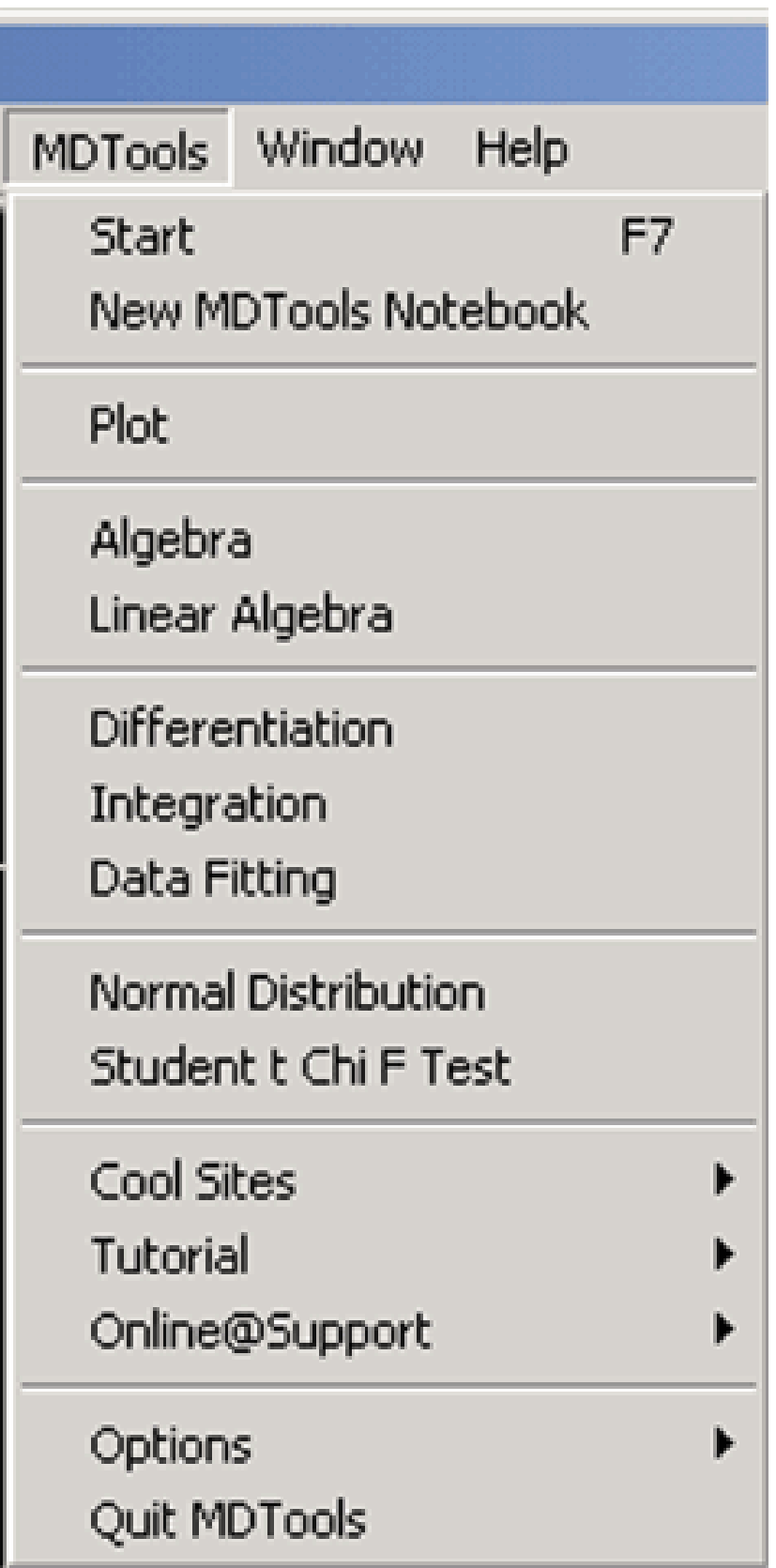}
\caption{MDTools installs its own menu entry in Mathematica}\label{MDT_Menu}
\end{minipage}
\hspace{0.3cm}
\begin{minipage}{11cm}
\includegraphics[width=11cm]{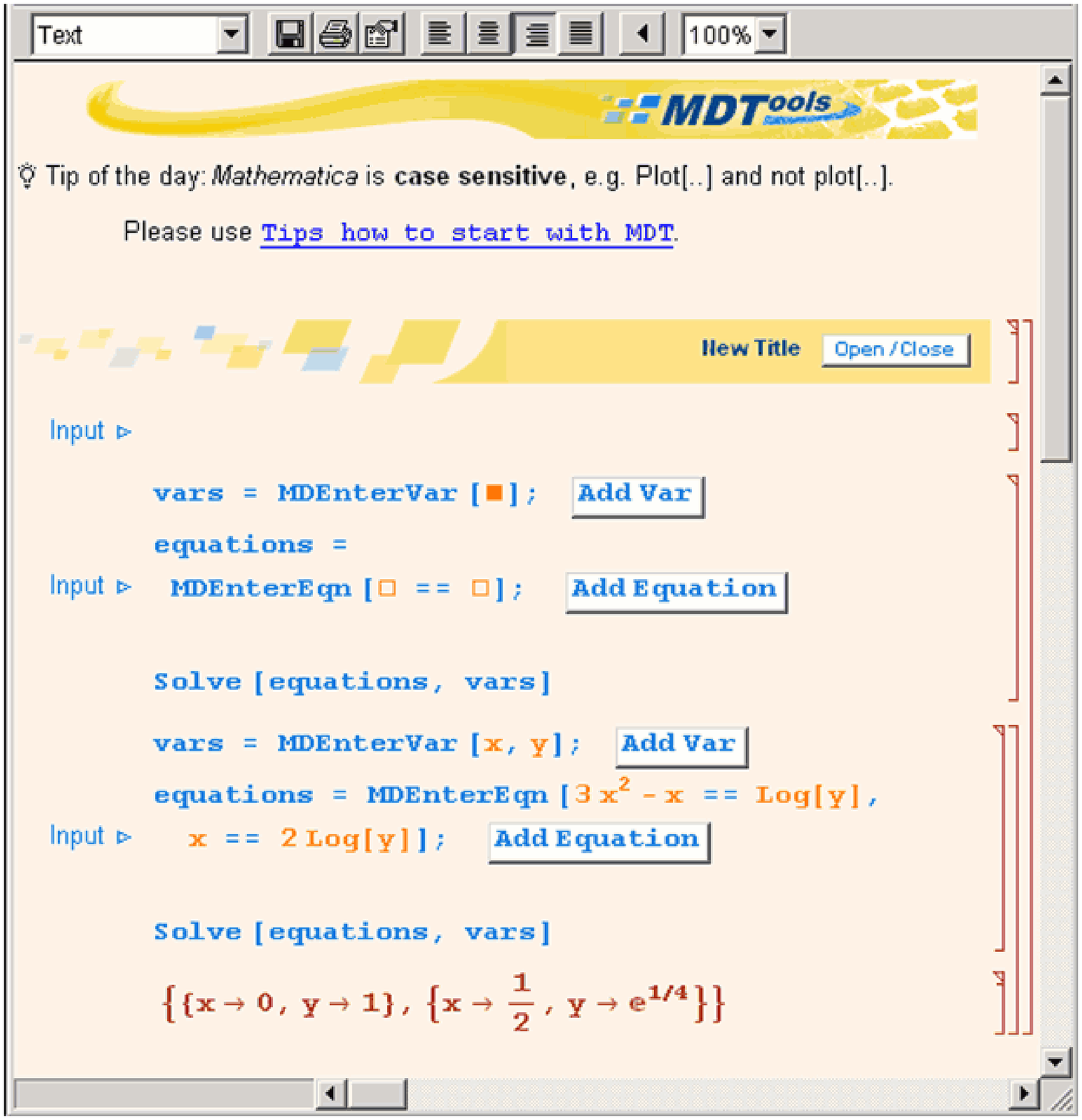}
\caption{A MDTools notebook with some templates for solving equations}\label{MDT_Notebook}
\end{minipage}
\end{figure}

M@th Desktop Tools is a set of helper palettes covering linear algebra, algebra, integration and statistics. The palettes provide templates  for the most important functions, which can be inserted directly into an active notebook by a simple click of the mouse. It can be installed using a setup program written entirely in Mathematica itself and sets up its own menu in Mathematica's menubar (see Fig.~\ref{MDT_Menu}) so that starting the program and opening the palettes is very straightforward.

The student works in a simple notebook (with an appealing look, but otherwise empty, Fig.~\ref{MDT_Notebook}) and solves the problems using the helper palettes. MDTools is not a full-feature e-learning tool but rather a utility for the Mathematica user which provides easy and quick access to its most important functions. The advantage over a pure Mathematica session is that, for one, the provided templates help the user avoid many syntactic errors, which are usually very time-consuming to track down without a good knowledge of Mathematica's expression syntax. Furthermore, since the templates provide all required fields, variables, and expression structures,  the user can concentrate on the important part of his or her work.

\subsection{Palettes}
\begin{figure}[t]\centering
\begin{minipage}{4cm}\centering
\includegraphics[width=3.5cm]{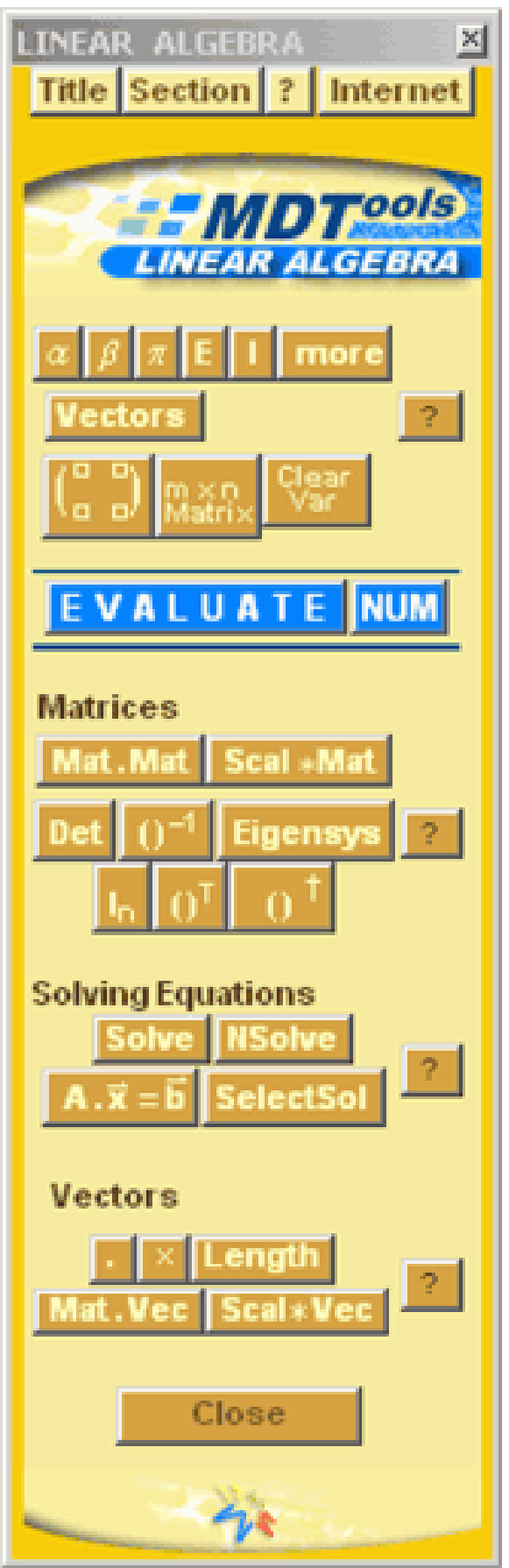}
\caption{MDTools' linear algebra palette}\label{MDT_Pal1}
\end{minipage}
\hspace{0.5cm}
\begin{minipage}{4cm}\centering
\includegraphics[width=3.5cm]{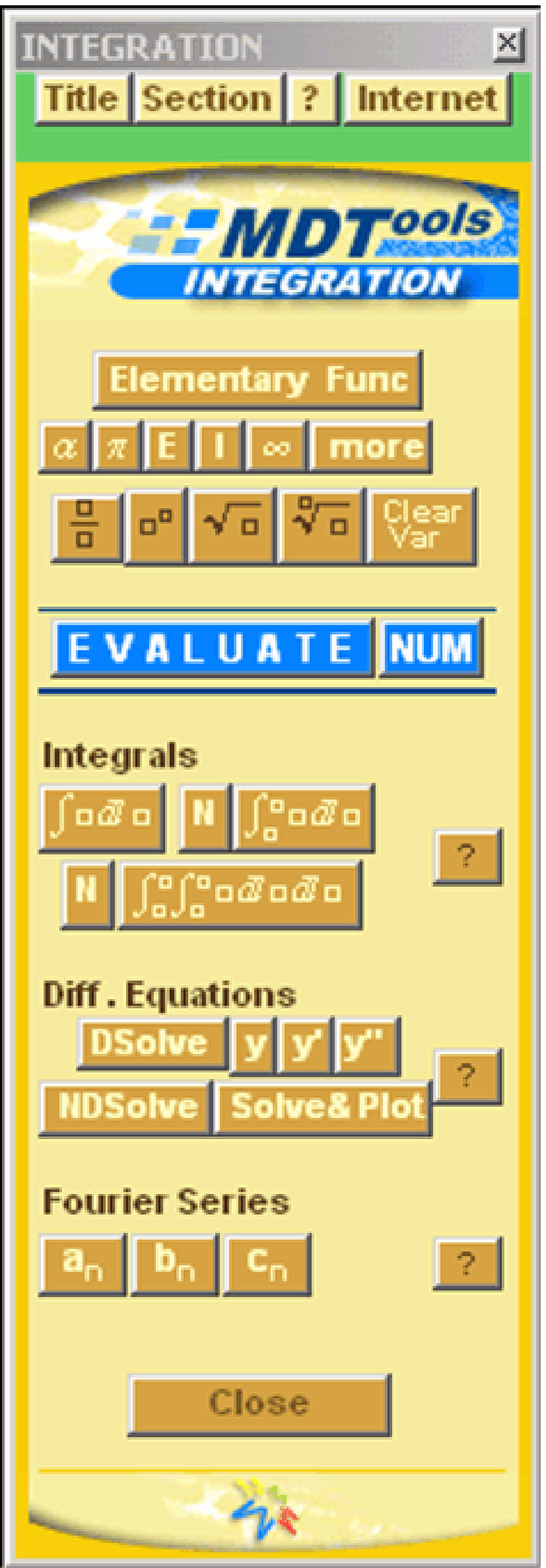}
\caption{MDTools' integration palette}\label{MDT_Pal2}
\end{minipage}
\hspace{0.5cm}
\begin{minipage}{4cm}\centering
\includegraphics[width=3.5cm]{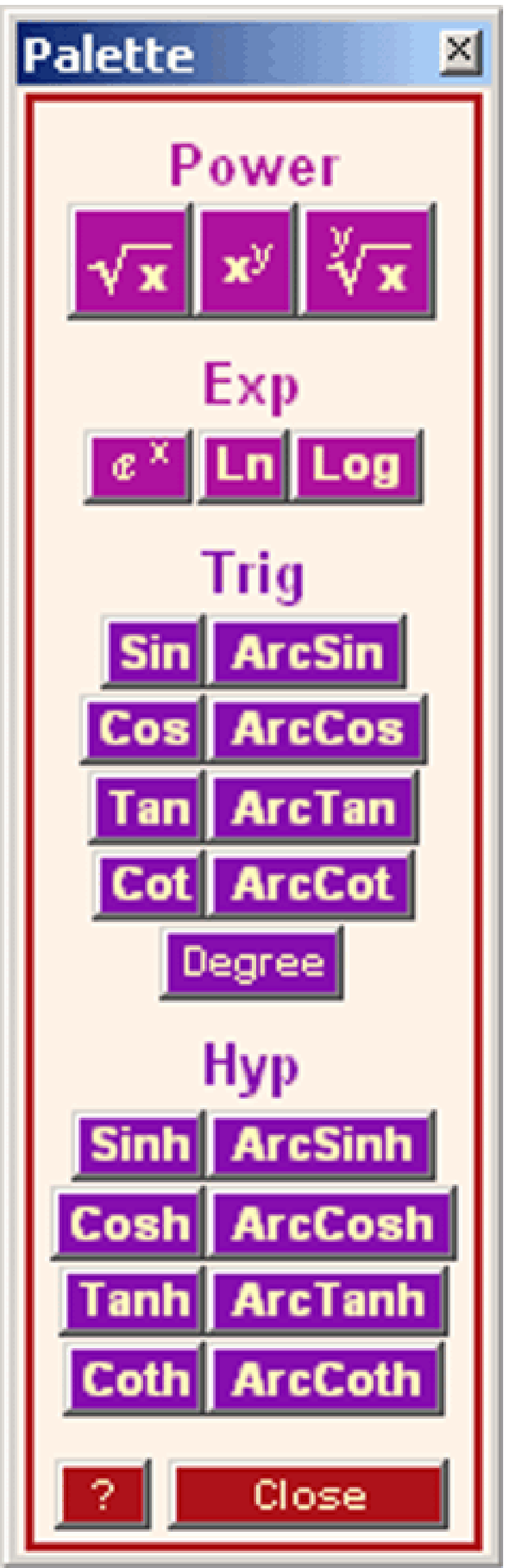}
\caption{MDTools' elementary functions helper palette}\label{MDT_Pal3}
\end{minipage}
\end{figure}
MDTools consists of eight main palettes, two of which are displayed in Fig.~\ref{MDT_Pal1} and \ref{MDT_Pal2}, covering the following topics:
\begin{itemize}
\item {\em Plot}: This palette provides easy access to several two- and three-dimensional plotting functions. 
\begin{itemize}
\item {\em Plot options helper palette}: The plot palette is supplemented by this palette, which covers the most frequently used options of the \texttt{*Plot} commands of Mathematica. The options are automatically inserted with the correct syntax at the appropriate position (not necessarily where the cursor stands) in the plot command, provided there is such a  command in the current cell of the active notebook. Furthermore, a good suggestion for the option value is given.
\end{itemize}
\item {\em Algebra}: This palette helps the user work with arbitrary (e.g. polynomial, trigonometric, etc.) expressions, such as simplifying or factoring terms.
\item {\em Linear Algebra}: As shown in Fig.~\ref{MDT_Pal1}, this palette supports the user in solving equations (linear, but also more general types) and working with vector and matrix calculus.
\item {\em Differentiation}: Calculations of full and partial derivatives, divergence, gradient, and series expansion of functions.
\item {\em Integration}: Integration, solving differential equations, and Fourier sequences (see Fig.~\ref{MDT_Pal1}).
\item {\em Data Fitting}: Loading and plotting lists of data, conducting linear and non-linear fits.
\item {\em Normal Distribution}: Probabilities, confidence intervals, and hypothesis tests using the Normal distribution.
\item {\em Student-t and $\chi^2$ Distribution}: Tests using the Student-t and $\chi^2$ distributions.
\end{itemize}
In addition to these main palettes, three helper palettes are provided:  The Plot Options helper palette which has already been discussed, the Elementary Functions palette, and the Special Symbols helper palette.

\subsection{Templates}
\begin{figure}[b]\centering
\includegraphics[width=8cm]{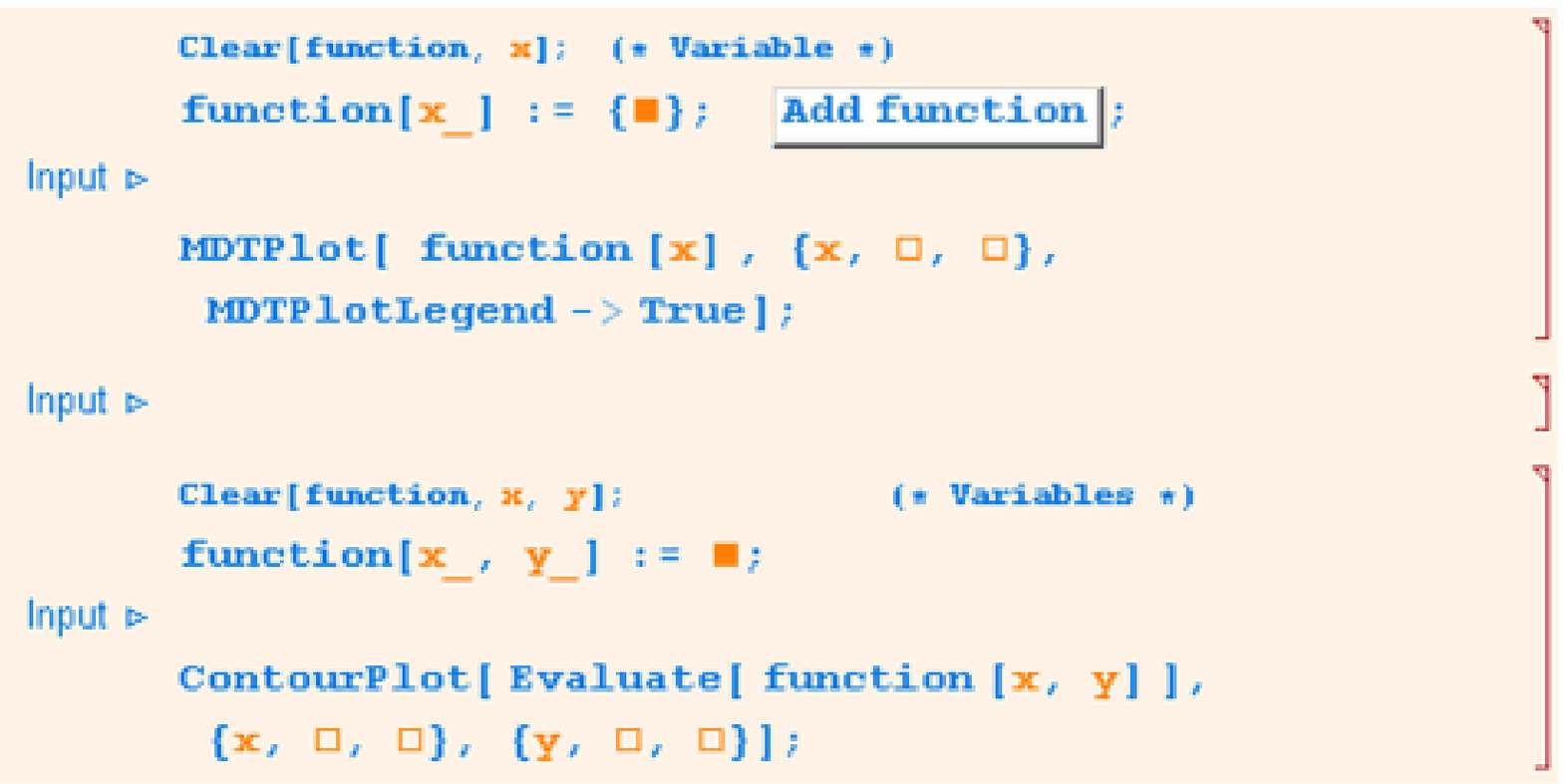}
\caption{Templates for plotting a function of one variable and creating a contour plot of a three-dimensional function}\label{MDT_Template1}
\end{figure}

\begin{figure}[b]\centering
\includegraphics[width=8cm]{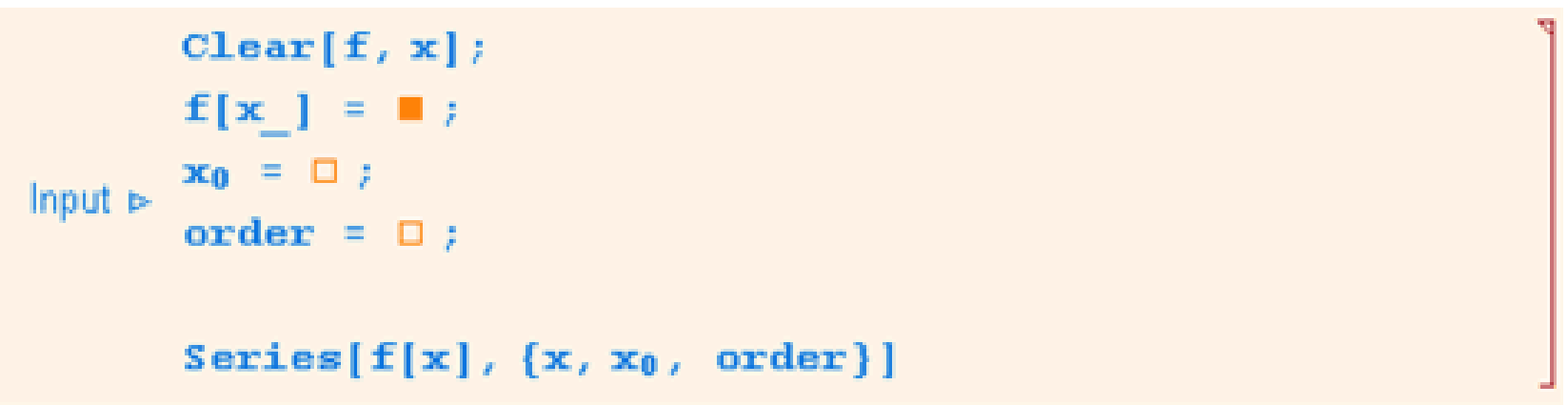}
\caption{Template for a Taylor series expansion up to a given order}\label{MDT_Template2}
\end{figure}

\begin{figure}[h]\centering
\includegraphics[width=7.8cm]{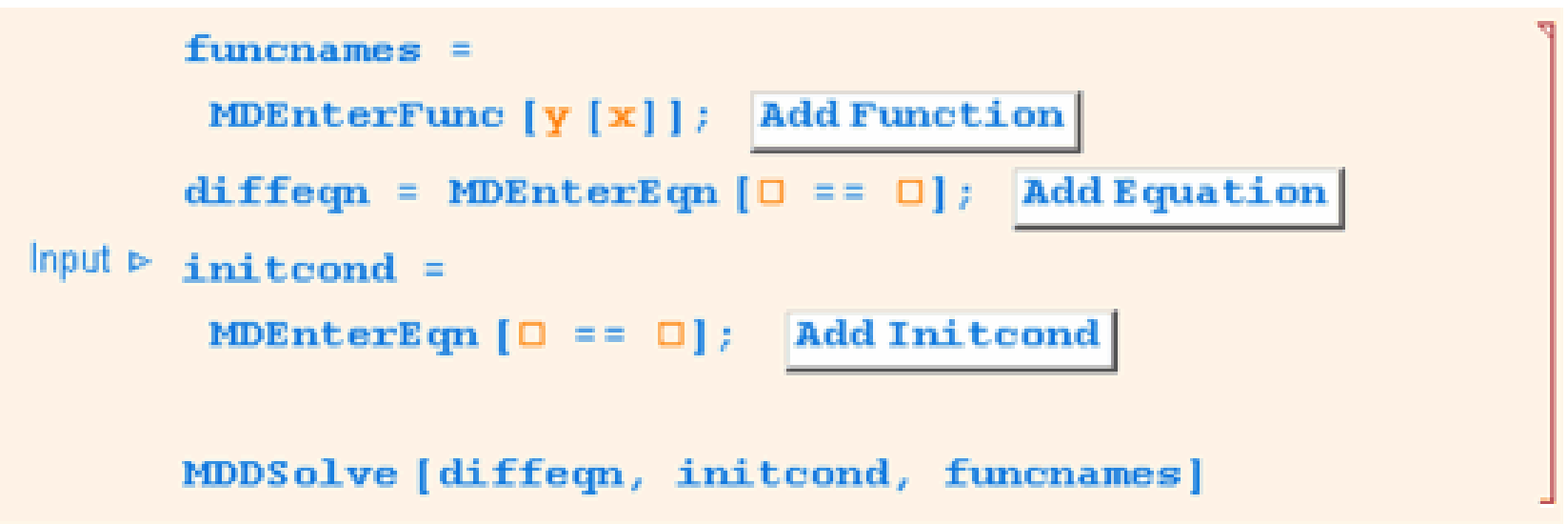}
\caption{Solving a differential equation with \texttt{DSolve}}\label{MDT_Template3}
\end{figure}

\begin{figure}[h]\centering
\includegraphics[width=7.8cm]{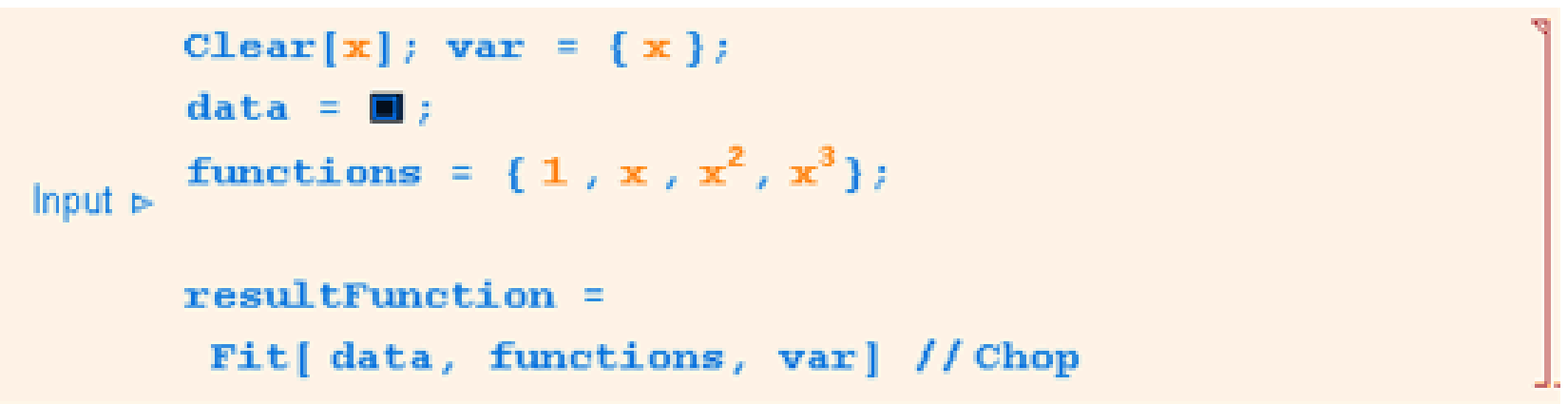}
\caption{Performing a least-squares fit of a polynomial to the given data}\label{MDT_Template5}
\end{figure}

All the palettes presented in the previous paragraph paste into the current notebook templates which already contain the correct syntax and provide placeholders for all input needed from the user. Some examples are given in Fig.~\ref{MDT_Template1} to \ref{MDT_Template5}, where one can see that the user may have to change only those parts colored in orange.

\begin{figure}[h]\centering
\includegraphics[width=7.8cm]{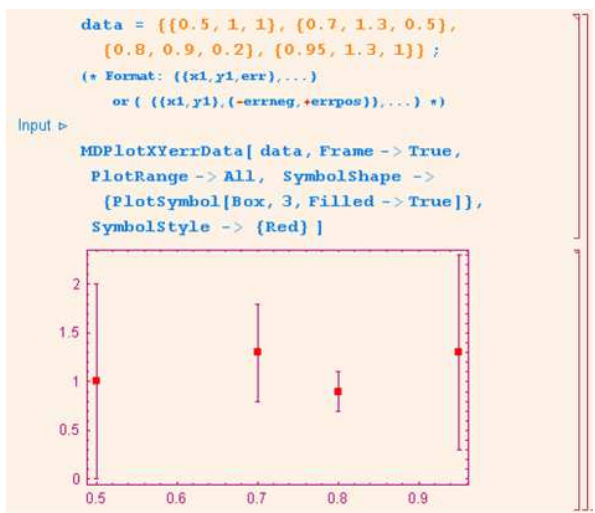}
\caption{Plotting two-dimensional data together with error bars.}\label{MDT_Template4}
\end{figure}

\section{M@th Desktop}\label{MD}
M@th Desktop, in contrast, is a full-featured e-learning and courseware package building on the extensive features of Mathematica. It currently provides separate modules for linear algebra, differentiation, integration, and statistics. Each of them is based on the same basic functionality and the same navigation and usage but can be installed separately.
\begin{figure}[p]\centering
\includegraphics[width=9cm]{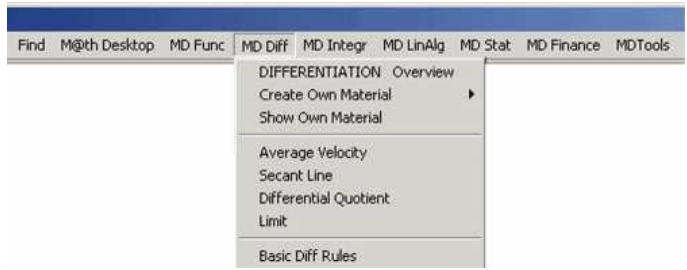}
\caption{MD installs its own menu items in Mathematica's menu.}\label{MD_Menu1}
\end{figure}
\begin{figure}[p]\centering
\includegraphics[width=13cm]{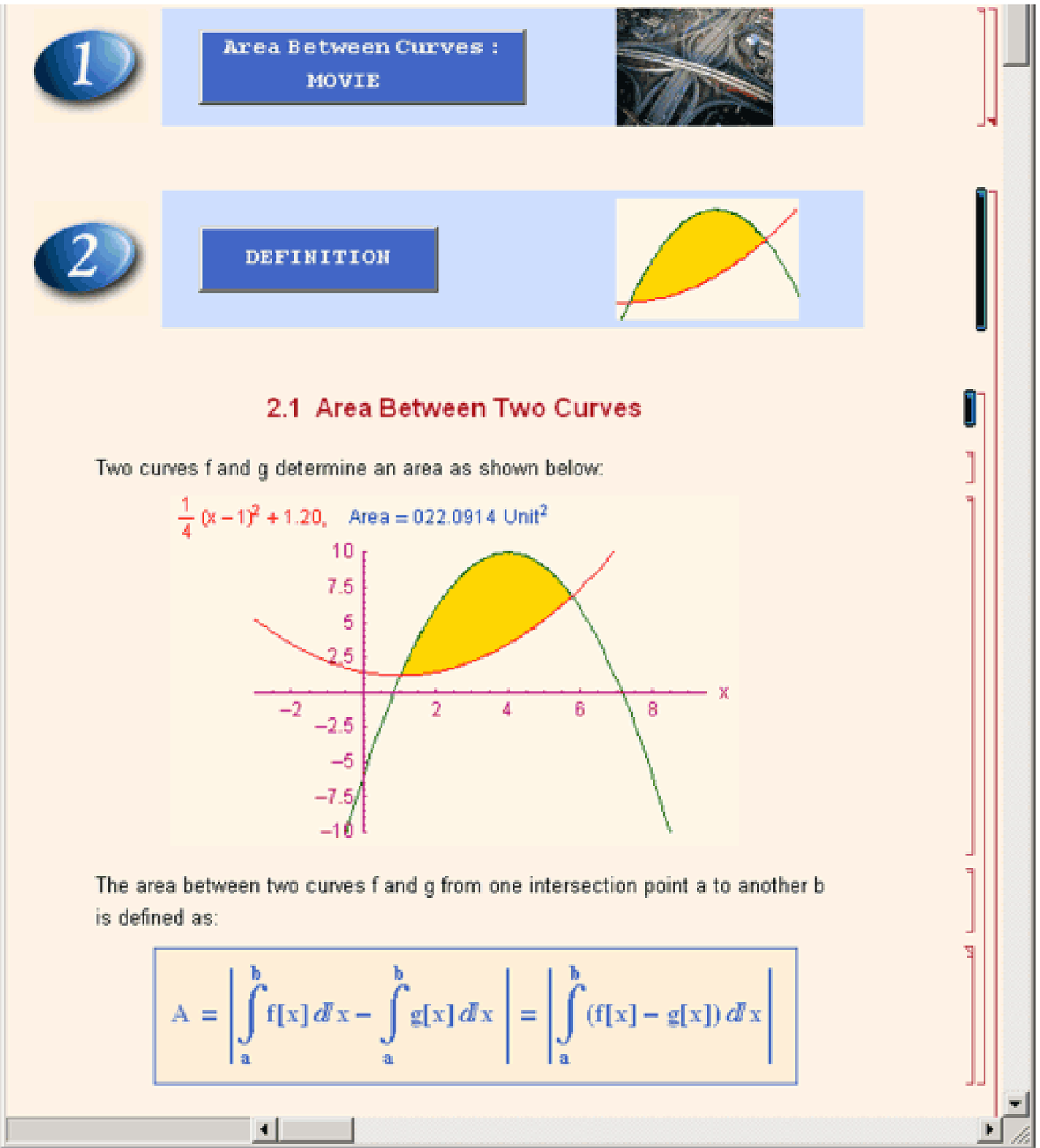}
\caption{Each notebook starts with the presentation of the mathematical foundations of the topic covered. The students need to understand these basic facts well before they can apply them later using the computer instead of manual calculations.}\label{MD_AreaBetweenCurves}
\end{figure}
Like M@th Desktop Tools, upon installation each module creates  its own menu item in Mathematica's menu bar (Fig.~\ref{MD_Menu1}). \vspace{0.5em}

Each of these modules, in turn covers all important topics of its subject (e.g. the integration module covers the definite  and indefinite integral, Riemann sums, area between curves, curve length, surfaces, volume, applications in physics, Laplace and Fourier Transformations and first-order differential equations), with each topic treated in its own one notebook and with an accompanying palette. 
\begin{figure}[bt]\centering
\includegraphics[width=13.5cm]{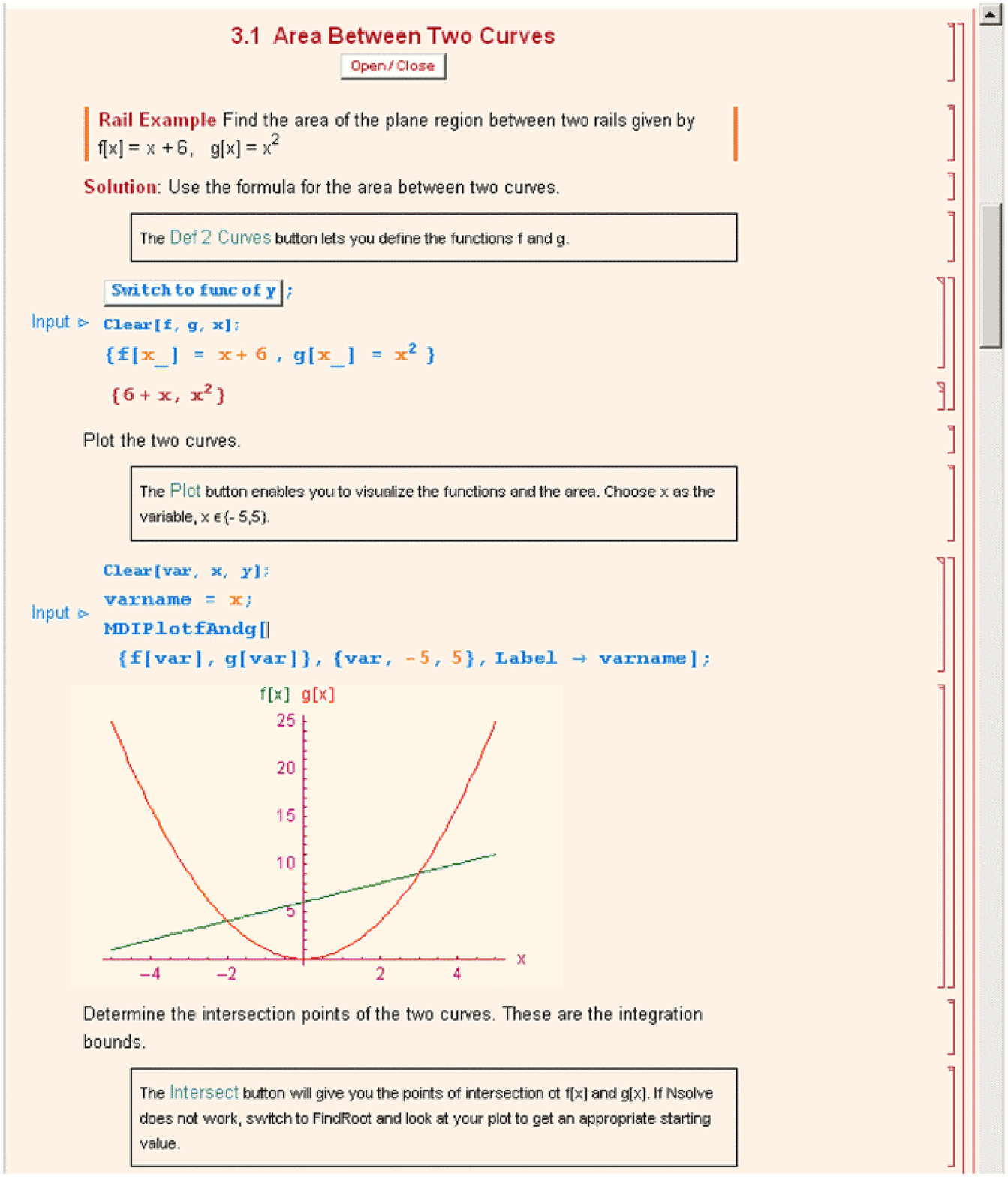}
\caption{When the students understand the principles, the computer can be used as a calculation tool to avoid lengthy manual calculations.}\label{MD_AreaBetweenCurves1}
\end{figure}

As an example, let us look at the problem of calculating the area between two curves which are functions of one variable. The notebook (Fig.~\ref{MD_AreaBetweenCurves}) starts with the mathematical and visual description of the problem. In this phase, the students are supposed to learn the principles and basic concepts which are needed for this particular type of problem. The teacher is also of enormous importance in this phase, because although the notebooks explain the problem and its solution in a basic way, they can never  completely replace the lecturer. 
The notebooks should be seen like a textbook,  which supports the teacher as far as notation and the statement of the basic principles goes. This phase in class is also the part where the students need to do several examples by hand. Buchberger \cite{Buchberger:1989} calls this the "white-box phase" in contrast to the later "black-box phase", when the computer is seen as a black box which simply takes the input and produces the result. By then, however, the students are already required to understand how the black box really works. Instead of reproducing the black-box algorithm again and again, they can concentrate on the more important step of interpreting given text problems and turning them into problems denoted by mathematical equations (Fig.~\ref{MD_AreaBetweenCurves1}).

\begin{figure}\centering
\includegraphics[width=13cm]{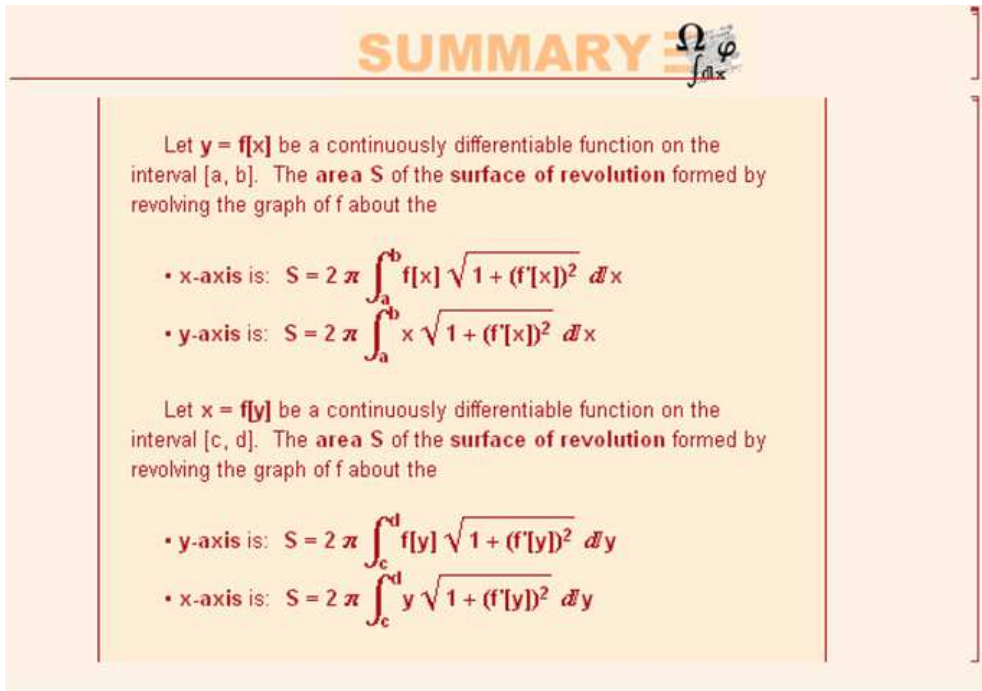}
\caption{Each notebook ends with a short summary of the underlying mathematical concepts.}\label{MD_SummarySurface}
\end{figure}

Finally, each notebook contains several sections with exercises (and their solutions so that the students can also check their results when they are studying on their own) and a short summary of the  basics presented in the notebook (e.g. Fig.~\ref{MD_SummarySurface} shows the summary of the notebook discussing the calculation of the surface of revolution).

\subsection{Notebooks, Palettes, and Helper Palettes}
\begin{figure}[bt]\centering
\includegraphics[width=12cm]{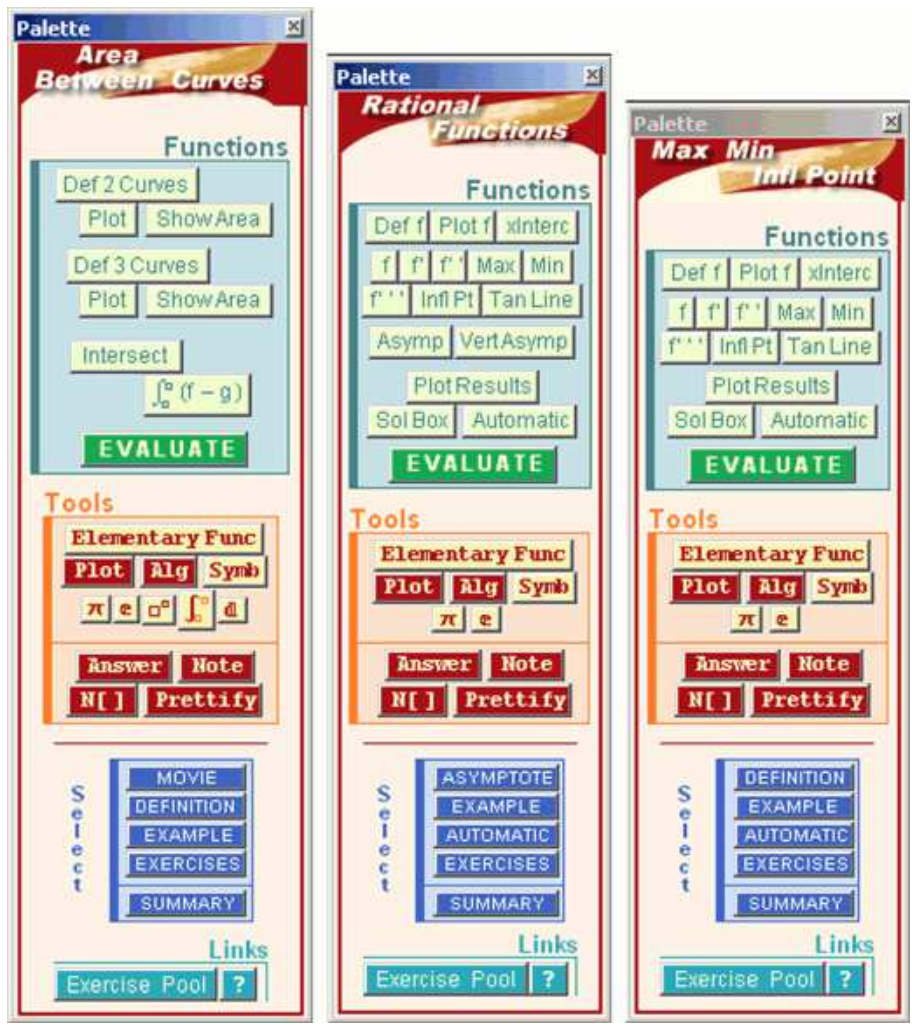}
\caption{Several palettes that accompany a M@th Desktop notebook covering one topic}\label{MD_Palettes}
\end{figure}

As described in the previous section, each topic is treated in one notebook and also accompanied by a palette specifically tailored to the current problem. Fig.~\ref{MD_Palettes} shows several of the palettes available in M@th Desktop.

The green "Functions" part on the top of the palette contains the templates for the most frequently used functions connected with the palette's topic, while the red "Tools" part in the middle of the palette is the same for all palettes and provides access to general features needed in all circumstances,  like plotting a function or solving equations. Also, it contains buttons for the insertion of pre-formatted cells for typing answers or short personal notes into the notebook. The final blue "Select" section of each palette links to the notebook dealing with the same topic and can be seen as a table of contents of the notebook. A click on any button jumps to the corresponding section of the linked notebook. 

\subsection{Preparing your own teaching material}
\begin{figure}[bt]\centering
\includegraphics[width=5.5cm]{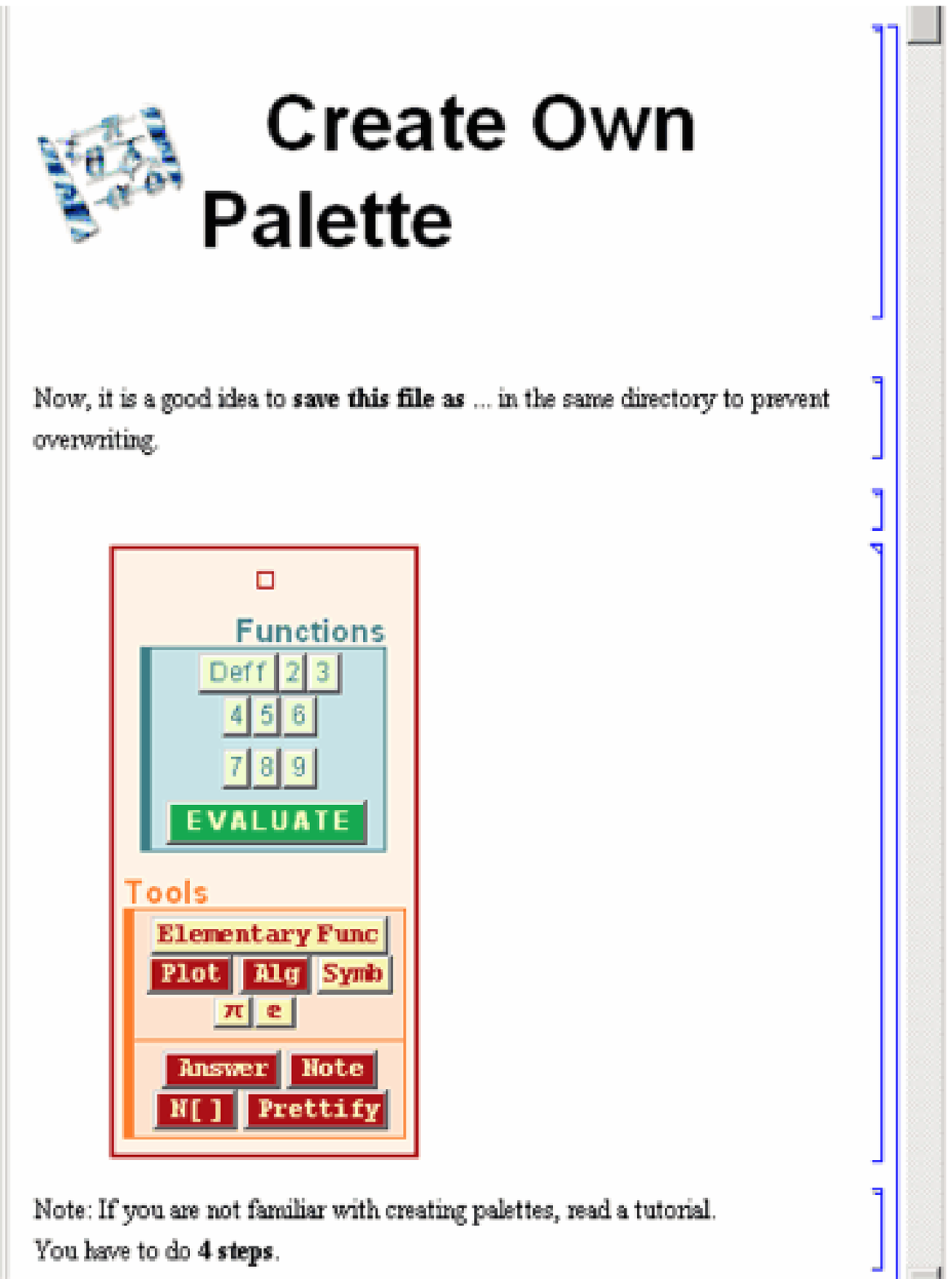}
\hspace{0.3cm}
\includegraphics[width=9.5cm]{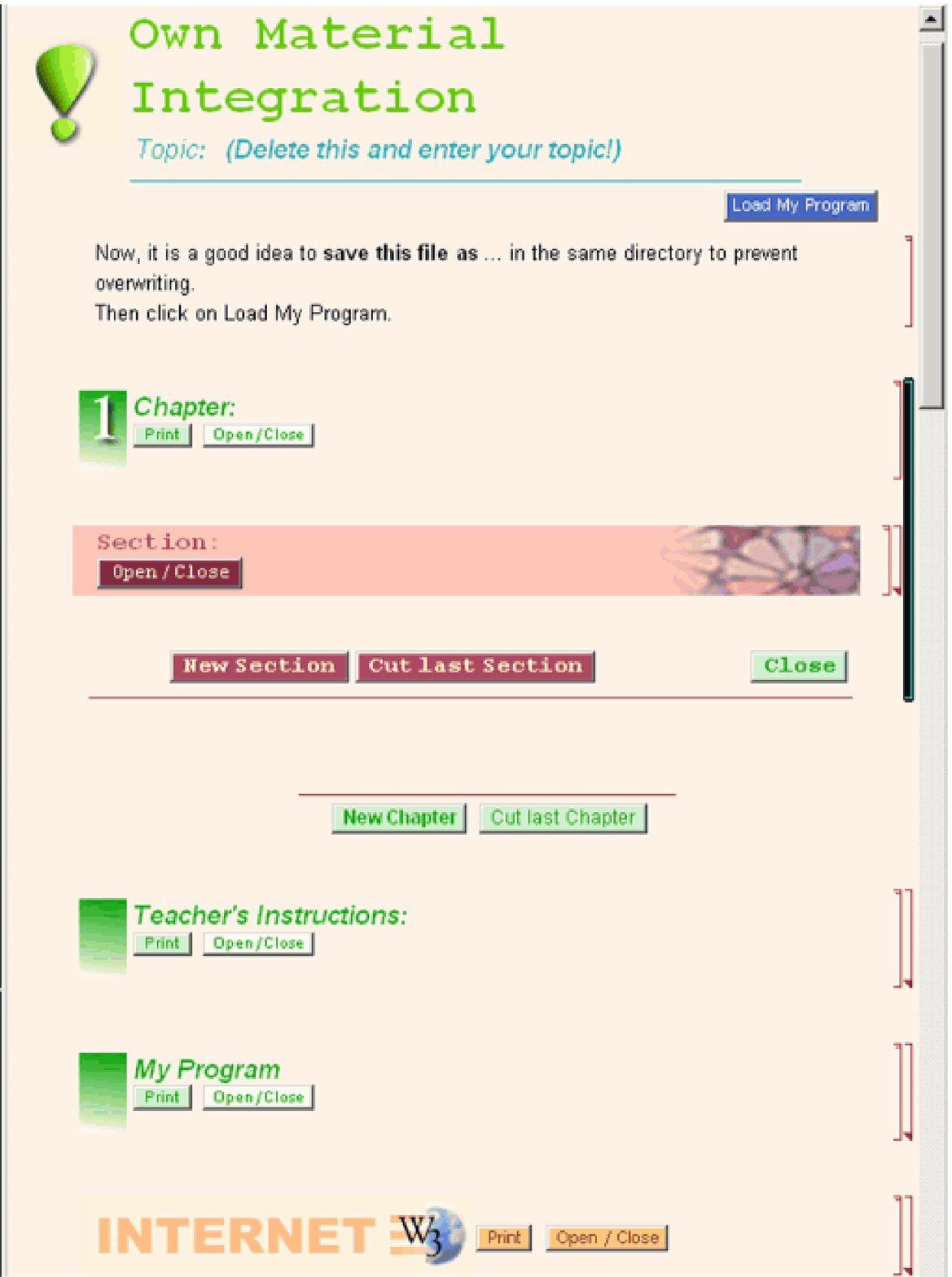}
\caption{Teachers can also create their own material (notebooks describing mathematical topics and palettes providing the supporting templates, functions, and navigation) in the M@th Desktop style and then integrate them into the MD system.}\label{MD_OwnMaterial}
\end{figure}

As a courseware package  for teaching mathematics, M@th Desktop also needs to give teachers and lecturers the possibility not only to create their own teaching material, but also to integrate it into the existing M@th Desktop system. For this reason, MD also provides features (Fig.~\ref{MD_OwnMaterial}) for creating your own notebook discussing specific topics, as well as for creating and accompanying palette.

\section{Conclusion}
In this paper we presented the two courseware packages M@th Desktop and M@th Desktop Tools based on Mathematica. Both of them are built on the paradigm of so-called "blended learning", in which computer and lecturer complement each other so that the students get the best of both conventional teaching and computer-based e-learning.

 While M@th Desktop Tools provides students with palettes and templates for the most important functions needed for their calculations, M@th Desktop is a full-featured modular courseware application with notebooks tackling specific topics. Also, helper palettes are provided for each notebook, which enable the students to quickly start doing their calculations and at the same time avoid the inescapable typing and syntax errors usually encountered when working with a computer algebra system which requires specific notation.

\nocite{Fuchs:1998}
\nocite{*}
\bibliographystyle{abbrv}
\bibliography{MathDesktop}

\noindent
{\em Reinhold Kainhofer } {\small (email: \url{kainhofer@deltasoft.at})}\\
{\em Reinhard V. Simonovits} {\small (email: \url{simonovits@deltasoft.at})}\\[0.5em]
{\small Deltasoft mathematics\\
Web: \url{http://www.deltasoft.at/}\\[0.5em]
Elisabethstra\ss{}e 42\\
A-8010 Graz\\
Austria
}

\end{document}